\documentclass[12pt]{article}

\usepackage{epsfig}
\usepackage{subfigure}
\def\singlespace
{\smallskipamount=3.75pt plus1pt minus1pt
\medskipamount=7.5pt plus2pt minus2pt
\bigskipamount=15pa plus4pa minus4pt \normalbaselineskip=12pt plus0pt
minus0pt \normallineskip=1pt \normallineskiplimit=0pt \jot=3.75pt
{\def\smallskip {\vskip\smallskipamount}} {\def\medskip
{\vskip\medskipamount}} {\def\bigskip {\vskip\bigskipamount}}
{\setbox\strutbox=\hbox{\vrule height10.5pt depth4.5pt width 0pt}}
\parskip 7.5pt \normalbaselines}

\def\middlespace
{\smallskipamount=5.625pt plus1.5pt minus1.5pt \medskipamount=11.25pt
plus3pt minus3pt \bigskipamount=22.5pt plus6pt minus6pt
\normalbaselineskip=22.5pt plus0pt minus0pt \normallineskip=1pt
\normallineskiplimit=0pt \jot=5.625pt {\def\smallskip
{\vskip\smallskipamount}} {\def\medskip {\vskip\medskipamount}}
{\def\bigskip {\vskip\bigskipamount}} {\setbox\strutbox=\hbox{\vrule
height15.75pt depth6.75pt width 0pt}} \parskip 11.25pt
\normalbaselines}

\def\doublespace
{\smallskipamount=7.5pt plus2pt minus2pt \medskipamount=15pt plus4pt
minus4pt \bigskipamount=30pt plus8pt minus8pt \normalbaselineskip=30pt
plus0pt minus0pt
\normallineskip=2pt \normallineskiplimit=0pt \jot=7.5pt
{\def\smallskip {\vskip\smallskipamount}} {\def\medskip
{\vskip\medskipamount}} {\def\bigskip {\vskip\bigskipamount}}
{\setbox\strutbox=\hbox{\vrule height21.0pt depth9.0pt width 0pt}}
\parskip 15.0pt \normalbaselines}

\newcommand{\be}{\begin{eqnarray}}
\newcommand{\ee}{\end{eqnarray}}

\def\ss2t{{s_{2 \theta}}}

\def\cc2t{{c_{2 \theta}}}

\begin{document}
\middlespace

\vskip 2cm
\begin{flushright} KEK-TH-1117\\
VEC/PHYSICS/P/1/2006-2007
\end{flushright}
\begin{center}
\Large 
{\bf A $3 \times 2$ texture for neutrino oscillations and leptogenesis} \\
\vskip 1cm 
Biswajoy Brahmachari 
\footnote{{\tt electronic address:~~biswajoy@sssc.in}} 
and 
Nobuchika Okada
\footnote{{\tt electronic address:~~okadan@post.kek.jp}}\\
\end{center}
\begin{flushleft}

$^1$ ~~Department of Physics, Vidyasagar Evening College\\
~~~~39, Sankar Ghosh Lane, Kolkata 700006, India \\
%
$^2$ ~~Theory Division, KEK\\
~~~~Tsukuba, Ibaraki 305-0801, Japan\\
\end{flushleft}

\begin{center}
\underbar{Abstract} \\
\end{center}

In an economical system with only two heavy right handed neutrinos, 
we postulate a new texture for $3 \times 2$  Dirac mass matrix $m_D$. 
This model implies one massless light neutrino and thus displays only 
two patterns of mass spectrum for light neutrinos, namely hierarchical 
or inverse-hierarchical. Both the cases can correctly reproduce all
the current neutrino oscillation data with a unique prediction
$m_{\nu_e \nu_e} = \frac{ \sqrt{\Delta m^2_{solar}}}{3} $ and 
$\sqrt{\Delta m^2_{atm}}$ for the hierarchical and the
inverse-hierarchical cases, respectively, which can be tested in next 
generation neutrino-less double beta decay experiments. Introducing a
single physical CP phase in $m_D$, we examine baryon asymmetry through 
leptogenesis. Interestingly, through the CP phase there are correlations 
between the amount of baryon asymmetry and neutrino oscillation
parameters. We find that for a fixed CP phase, the hierarchical case
also succeeds in generating the observed baryon asymmetry in our universe, 
plus a non-vanishing $U_{e3}$ which is accessible in future baseline
neutrino oscillation experiments.

\newpage

\section{Introduction} 
The origin of the observed baryon asymmetry in our universe, 
ratio of number of baryons to photons\cite{WMAP-3Y} 
\be
 \eta_B = \frac{n_B-n_{\bar{B}}}{n_\gamma} 
 = 6.1 \pm {0.2} \times 10^{-10} , 
\ee
is one of the major problems in cosmology. 
This number has been deduced from two independent observations. 
(1) From the existing abundance of light elements formed 
after big bang\cite{bau}.  
(2) Precision measurements of cosmic microwave 
background\cite{WMAP-3Y}.  

Leptogenesis\cite{LG1, LG2} may explain 
this observed asymmetry between matter and antimatter content 
of the universe. In explaining this asymmetry one first creates a
tiny lepton asymmetry in the early universe. This lepton asymmetry
is recycled into observed baryon asymmetry above the electroweak
scale via sphaleron interactions\cite{sph1}. This is possible since
sphaleron interactions remain in thermal equilibrium above
the electroweak scale, they violate $B+L$, and since they conserve $B-L$. 

It is widely believed that the lepton asymmetry is formed by 
out-of-equilibrium, lepton number violating, CP violating decay of 
heavy right handed neutrinos. Existence of heavy right handed neutrinos 
also give a natural framework for explaining smallness of neutrino mass via
see-saw\cite{seesaw} mechanism. If there are no symmetry structures
in the theory to make the right handed neutrinos stable, they must
decay. They have non-vanishing Yukawa couplings with Higgs scalars and
left handed doublets, complex in general, for them to do so. Therefore 
we study lepton asymmetry generated by the CP violating decays of heavy 
right handed neutrinos (with Majorana mass) at the early stage of our 
universe. 
Since leptogenesis involves no new interactions apart from those 
required for see-saw mechanism to succeed, we may expect that 
the Physics of neutrino oscillations would clarify some deep mystery 
of cosmology such as the observed asymmetry between matter and 
antimatter with which it is linked.

We know from pioneering works of Sakharov\cite{sakharov} that CP violation
is an essential ingredient for theories of matter-antimatter 
asymmetry. If there is just one right handed neutrino, the Dirac mass 
matrix is $3\times 1$ dimensional. Lepton fields can absorb all complex phases
and there is no source of CP violation. Therefore one fails to have
leptogenesis with just one right handed neutrino. If there are two
right handed neutrinos $\{N_1, N_2\}$, the Dirac mass matrix is 
$3 \times 2 $ dimensional. Let us choose a basis where the charged
lepton mass matrix as well as the heavy right handed neutrino mass
matrix are diagonal. In this case we cannot absorb all six complex
phases in the Dirac matrix plus two complex phases in the Majorana
matrix by redefining five lepton fields
$\{l_e,l_\mu,l_\tau,N_1,N_2\}$. 
After re-phasing, we have three physical CP phases 
 in the $3 \times 2$ Dirac mass matrix. 
Therefore at the minimum, two right handed neutrinos\cite{2rhn} 
are enough to bring in a CP violating decay 
and successful leptogenesis. 

Neutrino oscillations show that neutrinos have non-zero mass and that
there are off-diagonal entries in the mass matrix written in flavor basis. 
Solar and atmospheric neutrino oscillation experiments have explored 
neutrino masses and mixing patterns. 
Current best fit values are\cite{nuosc1,nuosc2,nuosc3},
\be 
&& 7.2 \times 10^{-5} < \Delta m^2_{12} < 9.2 \times 10^{-5} 
~~{\rm~eV^2},\nonumber \\
&& 1.4 \times 10^{-3} < \Delta m^2_{23} < 3.3 \times 10^{-3} 
~~{\rm~eV^2},\nonumber \\
&& 0.25 < \sin^2 \theta_{12} < 0.39, 
\nonumber \\
&& \sin^2 2 \theta_{23} > 0.9, 
\nonumber \\
&& |U_{e3}| < 0.22. 
\label{oscillation data}
\ee

Any model of leptogenesis is required to reproduce these masses and
mixing angles. 
It is indeed interesting to see that, via see-saw mechanism,  
existing neutrino data can give desired mass spectrum of heavy right 
handed neutrinos plus right magnitudes of primordial lepton asymmetry. 
There are many studies of this kind where there are three
heavy right handed neutrinos and generated lepton asymmetry
depends on the form of the Yukawa texture\cite{mike}.
In this paper, we examine the system with only two right handed neutrinos. 
As discussed above, this system is the minimum one 
 to bring physical CP phases in the lepton sector. 
Number of free parameters in the neutrino sector 
 is much reduced compared to the usual three right handed neutrino case. 
However the system still contains an enough number of free parameters 
 to reproduce the current neutrino oscillation data. 
We introduce a texture for 3 $\times$ 2 Dirac neutrino mass matrix 
 by which the number of free parameters is further reduced. 
With a small number of free parameters, 
 we investigate neutrino oscillation parameters and 
 the amount of baryon asymmetry through leptogenesis. 
We will see correlations between them through a CP phase.

This paper is organized as follows. 
In the next section, we introduce $3 \times 2$ Dirac mass matrix 
 and a texture for it. 
We begin with the CP invariant case and apply a simple ansatz 
 to the light neutrino mass matrix so as to reproduce 
 the current best fit values for the neutrino oscillation parameters.
In Sec. 3, we introduce a single CP phase and examine 
 baryon asymmetry generated through leptogenesis 
 and neutrino oscillations. 
With four input parameters 
 (two light and two heavy neutrino mass eigenvalues), 
 all neutrino oscillation parameters as well as the baryon asymmetry 
 through leptogenesis are shown as a function of only the CP phase. 
Also, the averaged neutrino mass relevant to neutrino-less double 
 beta decay experiments and the Jarlskog invariant characterizing 
 CP violation in the lepton sector are presented 
 as a function of the CP phase. 
We see correlations among these outputs, and 
 presents some predictions for a fixed CP phase. 
The last section is devoted to conclusions. 

%
%

\section{Texture and a simple ansatz in CP invariant case}

Without loss generality, we begin with a reference basis 
in which the charged lepton mass matrix $m_l$, Dirac mass matrix $m_D$ 
and the right handed Majorana mass matrix $M_R$ are written as 
\be 
m_l=\pmatrix{m_e &0 &0\cr 0 & m_\mu & 0 \cr 0 & 0 & m_\tau},~~
m_D=\pmatrix{c_1 e^{i \delta_1}  & c_2 e^{i \delta_2} 
& c_3 e^{i \delta_3} \cr c_4 & c_5 & c_6},~~
M_R=\pmatrix{M_1 & 0 \cr 0 & M_2},  
\ee
where all parameters are real and $0 < M_1 < M_2$. 
We have three physical CP phases in $m_D$ by re-phasing. 
There is no triplet Higgs in the model, so left handed neutrinos 
do not have a Majorana mass at the beginning. 

After see-saw mechanism, the light neutrino mass matrix becomes 
\be 
m_\nu=m^T_D M^{-1}_R m_D, \label{seesaw} . 
\ee
Note that the system with $3 \times 2$ Dirac mass matrix 
 leads to 
\be 
\mbox{Det} (m_\nu)=0.
\ee
Therefore, at least, one mass eigenvalue of light neutrinos is zero. 
Concerning the current best fit values of neutrino oscillation data, 
we can conclude that only two patterns of diagonalized mass matrix 
for light neutrinos are possible. 
One is the so-called hierarchical case, 
\be 
 D_\nu = diag(m_1=0, m_2, m_3) , 
\ee  
with $m_2 = \sqrt{\Delta m_{12}^2}$ and 
$m_3 = \sqrt{\Delta m_{12}^2 + \Delta m_{23}^2}$. 
The other is the so-called inverse-hierarchical case, 
\be 
 D_\nu = diag(m_1, m_2, m_3=0) , 
\ee  
with $m_1 = \sqrt{ - \Delta m_{12}^2 + \Delta m_{23}^2}$ 
and  $m_2 = \sqrt{\Delta m_{23}^2}$.

Now we introduce a texture for the Dirac mass matrix 
 as $c_1=0$ and $\delta_2=0$, and $m_D$ becomes a more simple form, 
\be
m_D=\pmatrix{0  & c_2 & c_3 e^{-i \phi} \cr c_4 & c_5 & c_6},
\label{texture}
\ee
with a single CP phase $\phi$. 
The texture reduces the number of free parameters 
 into six and allows us to analyze the correlations 
 between the amount of baryon asymmetry and 
 neutrino oscillation parameters 
 with the single CP-phase. 
Similar textures have been discussed in Ref.~\cite{2rhn}. 
The explicit form of the light neutrino mass matrix 
 is given by 
\be
m_\nu= m_D^T M_R^{-1} m_D = 
 \pmatrix{
 {c_4^2 \over M_2}  & {c_4 c_5 \over M_2} & 
 {c_4 c_6  \over M_2} \cr 
 {c_4 c_5 \over M_2} & {c_2^2 \over M_1} + {c_5^2 \over M_2} &
 {c_2 c_3 \over M_1}e^{-i \phi} + {c_5 c_6  \over M_2}  \cr
 {c_4 c_6  \over M_2} & 
 {c_2 c_3 \over M_1}e^{-i \phi} + {c_5 c_6  \over M_2}  &
 {c_3^2 \over M_1}e^{-2 i \phi} + {c_6^2 \over M_2}  
}.
\label{mass_nu} 
\ee
Six parameters in the Dirac mass matrix and two heavy neutrino masses, 
eight parameters in total, correspond to physics of neutrinos.

We first tackle only neutrino oscillations in the CP invariant case, 
 $\phi=0$. 
CP violation will be introduced in the next section.  
As our stating point, we impose a simple ansatz that $m_\nu$ 
 is diagonalized by the so-called tri-bimaximal 
 mixing matrix\cite{hps}, 
\be 
D_\nu =U_{TB}^T m_\nu U_{TB} ~~~{\rm where}~~~ 
U_{TB}=\pmatrix{\sqrt{2 \over 3}
& \sqrt{ 1 \over 3} & 0 \cr -\sqrt{1 \over 6} & \sqrt{ 1 \over 3} &
\sqrt{ 1 \over 2} \cr -\sqrt{1 \over 6} & \sqrt{ 1 \over 3} &
-\sqrt{1 \over 2}}.
\label{ansatz}
\ee
In fact, the tri-bimaximal mixing matrix is in excellent agreement 
with the current best fit values in Eq.~(\ref{oscillation data}). 
This ansatz strongly constrains the parameters in Eq.~(\ref{texture}). 
For the hierarchical case, 
%
%
solving Eq.~(\ref{ansatz}) with Eq.~(\ref{mass_nu}), 
we can describe each component in the Dirac mass matrix 
 in terms of $m_2$, $m_3$, $M_1$ and $M_2$, 
%
%
%
\be
&&c_2 = - c_3 = \sqrt{M_1 m_3 \over 2}, \nonumber \\ 
&&c_4 = c_5=c_6= \sqrt{M_2 m_2 \over 3}. 
\label{H-comp}
\ee
For the inverse-hierarchical case, 
we find 
%
%
\be
&&c_2= c_3 =\sqrt{3 M_1 m_1 m_2 \over 2 (2 m_1+m_2)}, \nonumber \\ 
&&c_4= \sqrt{ M_2( 2 m_1+ m_2) \over 3}, \; 
c_5=c_6= (-m_1+m_2) \sqrt{ M_2 \over 3 (2 m_1+m_2) },  
\label{IH-comp}
\ee
in terms of $m_1$, $m_2$, $M_1$ and $M_2$. 
Thanks to our ansatz, only four parameters, 
 $m_1$ (or $m_3$), $m_2$, $M_1$ and $M_2$ are left free. 
These parameters will be used as inputs 
 for our analysis in the next section.

Now we can discuss experimental tests of the model. 
Neutrino-less double beta decay experiments give upper bounds 
on the averaged neutrino mass, which can be extracted 
from the $\nu_e \nu_e$ element of the Majorana mass matrix 
in the flavor basis. 
Here we see that for the hierarchical case, 
\be 
m_{\nu_e \nu_e}= (U_{TB} D_\nu U_{TB}^T)_{11} = 
 {m_2 \over 3} = {\sqrt{\Delta m^2_{12}} \over 3},  
\label{dbetaH}
\ee
while for the inverse-hierarchical case, 
\be 
m_{\nu_e \nu_e}= (U_{TB} D_\nu U_{TB}^T)_{11} = 
{2 m_1+ m_2 \over 3} 
 \simeq {\sqrt{\Delta m^2_{23}}}. 
\label{dbetaIH}
\ee 
Here we have used an approximation 
$m_1 \simeq m_2 \simeq \sqrt{\Delta m^2_{23}}$ 
for $\Delta m^2_{12} \ll \Delta m^2_{23}$, 
in the inverse-hierarchical case. 
Therefore we have a chance of testing this relations 
in future neutrino-less double beta decay experiments 
with the sensitivity $m_{\nu_e \nu_e} \geq 10^{-3}$ eV.

\section{Numerical analysis in CP violating case} 

In this section we will introduce non-zero CP phase $\phi$ 
 in the texture of Eq.~(\ref{texture}). 
Except the CP phase, parameters in the Dirac mass matrix 
 are described in terms of four free parameters 
 as in Eq.~(\ref{H-comp}) or in Eq.~(\ref{IH-comp}). 
With non-zero CP phase, we can obtain baryon asymmetry 
 through leptogenesis. 
In addition, the Dirac mass matrix becomes complex and 
 resultant neutrino oscillation parameters are 
 deviating from those in the CP invariant case. 
We will see correlations among resultant neutrino oscillation 
 parameters and the amount of baryon asymmetry created 
 via leptogenesis.

Let us first consider leptogenesis. 
Primordial lepton asymmetry in the universe is generated 
 through CP violating out-of-equilibrium decay of the lightest 
 heavy neutrinos, which is characterized by 
 the CP violating parameter $\epsilon$\cite{vissani}, 
\be
\epsilon = - {3 \over 4 \pi v^2} {1 \over [m_D m_D^\dagger]_{11}}
{\rm Im}[(m_D m_D^\dagger)_{12}^2] 
 \; F \left( {M^2_2 \over M^2_1} \right) .
\label{eps}
\ee
This formula for asymmetry is valid in the basis where the
right handed neutrino is diagonal. Here, $v=246$ GeV is the 
vacuum expectation value of Higgs field, 
 and $F(x)=\sqrt{x}~ [{2 \over x-1} + \ln [{1+x \over x} ]]$ 
 and $F(x) \simeq 3/\sqrt{x}$ for $ x \gg 1$. 
Sphaleron processes will convert this lepton asymmetry 
 into baryon asymmetry and, as a result, 
 the baryon asymmetry is approximately described as
\be
\eta_B = 0.96 \times 10^{-2}~(-\epsilon)~\kappa. 
\label{etaB}
\ee
Here $\kappa < 1 $ is the efficiency factor, 
 that parameterizes dilution effects for generated lepton asymmetry 
 through washing-out processes. 
To evaluate the baryon asymmetry precisely, 
 numerical calculations \cite{lgn1} are necessary. 
We use a fitting formula of the efficiency factor 
 given in terms of effective light neutrino mass $\tilde{m}$ 
 such that\cite{lgn2}
\be
\kappa = 2  \times 10^{-2} 
 \left( 
 { 0.01 \mbox{eV} \over \tilde{m}} \right)^{1.1},
~~\mbox{where}~~ \tilde{m} = {(m_D m_D^\dagger)_{11} \over M_1} .  
\label{kappa}
\ee

Using the above formulas, we estimate the baryon asymmetry 
 as a function of only the single CP phase $\phi$ 
 with inputs $m_1$ (or $m_3$), $m_2$, $M_1$ and $M_2$. 
Numerical results are shown in Fig.~1 
 for the hierarchical and the inverse-hierarchical cases, 
 respectively. 
Here we have taken 
 $m_2= 9.59 \times 10^{-3}$ eV, 
 $m_3= 4.56 \times 10^{-2}$ eV, 
 $M_1=10^{13}$ GeV and $M_2=10^{14}$ GeV 
 for the hierarchical case, 
 while 
 $m_1= 4.46 \times 10^{-3}$ eV, 
 $m_2= 4.56 \times 10^{-2}$ eV, 
 $M_1=10^{13}$ GeV and $M_2=10^{14}$ GeV 
 for the inverse-hierarchical case. 
We can see that in the hierarchical case, 
 $\phi= 0.668$ or $3.075$(rad) provides the baryon asymmetry 
 consistent with the current observations. 
On the other hand, the inverse-hierarchical case cannot 
 provide sufficient baryon asymmetry.

To understand these results, it is useful to give explicit formulas  
 for leptogenesis 
 in terms of parameters in the texture of Eq.~(\ref{texture}). 
The CP violating parameter and the effective mass $\tilde{m}$ 
 are, respectively, written as 
\be
 \epsilon &=& 
 \frac{1}{2 \pi v^2 } 
 \frac{c_3 c_6 (c_2 c_4 + c_3 c_6 \cos \phi) \sin \phi}{c_2^2+c_3^2}, 
\nonumber \\
\tilde{m} &=& \frac{c_2^2+c_3^2} {M_1}.  
\ee
In the hierarchical case, the parameters fixed in Eq.~(\ref{H-comp}) 
 gives 
\be
\eta_B \simeq 2.5 \times 10^{-8} 
  \left( \frac{M_1}{10^{13} \mbox{GeV}}  \right)   
 \left( \frac{m_2}{0.01 \mbox{eV}}  \right) 
  \left( \frac{0.01 \mbox{eV}}{m_3}  \right)^{1.1}  
  (1-\cos \phi) \sin \phi . 
\ee
Here we have used an approximation formula 
 $F(M_2^2/M_1^2) \simeq 3 M_1/M_2$, assuming $M_1 \ll M_2$. 
Our result is independent of $M_2$ as long as $M_1 \ll M_2$. 
To obtain a formula for the inverse-hierarchical case, 
 we use an approximation, 
 $m_1=\sqrt{-\Delta m_{12}^2 + \Delta m_{23}^2} \simeq 
 \sqrt{\Delta m_{23}^2} ( 1+ 0.5 (\Delta m_{12}^2/\Delta m_{23}^2) )$. 
Thus the parameters in Eq.~(\ref{IH-comp}) lead to 
\be
\eta_B \simeq -2.1 \times 10^{-9} 
  \left( \frac{M_1}{10^{13} \mbox{GeV}}  \right) 
 \left( \frac{m_2}{0.01 \mbox{eV}}  \right) 
 \left( \frac{\Delta m_{12}^2}{\Delta m_{23}^2}  \right)^2 
 \left( \frac{0.01 \mbox{eV}}{m_2}  \right)^{1.1} 
  (1-\cos \phi) \sin \phi . 
\ee
Here we have again used $F(M_2^2/M_1^2) \simeq 3 M_1/M_2$, 
 assuming $M_1 \ll M_2$, and the result is independent of $M_2$. 
The baryon asymmetry is suppressed by the factor, 
 $\left( \Delta m_{12}^2/\Delta m_{23}^2 \right)^2$. 
If $M_1$ is very large, for example $M_1 \geq 10^{15}$ GeV, 
 we can give sufficient baryon asymmetry even with the suppression. 
However, in thermal leptogenesis re-heating temperature 
 after inflation would be larger than the lightest heavy neutrino mass. 
It would be difficult to achieve such a quite high reheating temperature 
 in usual reheating scenarios. 
We need some other mechanism 
 such as a resonant leptogenesis\cite{rlg} 
 to enhance the primordial lepton asymmetry.

Now we analyze the neutrino oscillation parameters 
 in the case of non-zero CP phase. 
Parameters $c_i$ in the texture are fixed as discussed in 
 the previous section, and lead to the tri-bimaximal mixing matrix 
 in the CP invariant case. 
When we switch CP phase on, the Dirac mass matrix becomes complex 
 and, as a result, output oscillation parameters are deviating 
 from the CP invariant case. 
In particular, we will find non-vanishing $U_{e3}$.

Substituting parameters given in Eq.~(\ref{H-comp}) or Eq.~(\ref{IH-comp}) 
 into the light neutrino mass matrix of Eq.~(\ref{mass_nu}), 
 we find that $m_\nu$ is independent of $M_1$ and $M_2$ 
 even for non-zero CP phase. 
Therefore, with input parameters $m_1$ (or $m_3$) and $m_2$, 
 output oscillation parameters are functions of 
 only the CP phase $\phi$.

In the hierarchical case with inputs 
 $m_2=9.59 \times 10^{-3}$ eV and $m_3=4.56 \times 10^{-2}$ eV, 
 resultant oscillation parameters are depicted in Fig.~2-4. 
For CP phase $\phi \leq 0.898$(rad), 
 outputs corresponding to the solar neutrino oscillation 
 are consistent with the best fit values, 
 while other outputs are within the best fit region 
 for any values of CP phase. 
For $\phi=0.668$(rad), 
 which provides the observed baryon asymmetry 
 $\eta_b = 6.1 \times 10^{-10}$, 
 we find the following neutrino oscillation parameters: 
\be
\Delta m^2_{12}&=& 8.1 \times 10^{-5}~~ {\rm eV^2}, 
 \nonumber \\ 
\Delta m^2_{23}&=& 2.0 \times 10^{-3}~~ {\rm eV^2},
 \nonumber \\
\sin^2 \theta_{12} &=& 0.36 , 
 \nonumber \\
\sin^2 2 \theta_{23} &=& 1.0,  
  \nonumber \\ 
|U_{e3}| &=& 0.029 . 
\ee
They are all consistent with observations. 
Non-vanishing $U_{e3}$ is our prediction, 
 whose value would be covered in future baseline 
 neutrino oscillation experiments. 
As can be seem from Eq.~(\ref{mass_nu}), 
 the $\nu_e\nu_e$ element of $m_\nu$ is independent of 
 the CP phase and we obtain the same result as Eq.~(\ref{dbetaH}), 
 numerically, 
\be
m_{\nu_e \nu_e}=  3.2 \times 10^{-3} \; 
 \mbox{eV}. 
\ee 

In the inverse-hierarchical case with input parameters 
 $m_1=4.46 \times 10^{-3}$ eV and $m_2=4.56 \times 10^{-2}$ eV, 
 resultant oscillation parameters are depicted in Fig.~5 and 6. 
In this case, we find 
 $\Delta m_{23}^2 \simeq 2.05 \times 10^{-3}$ eV$^2$ 
 and $\sin^2 \theta_{23}=1$, (almost) independent of the CP-phase. 
Although the inverse-hierarchical case cannot provide 
 the observed baryon asymmetry, 
 output oscillation parameters are consistent 
 with the current data for a small CP phase $\phi \leq 1.04$(rad). 
Again, the $\nu_e\nu_e$ element of $m_\nu$ 
 is independent of the CP phase, 
 and we obtain the same result as Eq.~(\ref{dbetaIH}), 
 numerically, 
\be 
m_{\nu_e \nu_e}=  4.5 \times 10^{-2} \; 
 \mbox{eV}. 
\ee 
This is an order of magnitude larger than the value 
 in the hierarchical case.

It is also interesting to see a correlation between 
 the baryon asymmetry through leptogenesis and 
 the leptonic CP violating phase (Dirac phase)\cite{PPR}. 
CP violation in the lepton sector is characterized 
 by the Jarlskog invariant\cite{Jarlskog}, 
\be 
 J_{CP}&=&\mbox{Im}[U_{e2} U^{*}_{\mu 2} U^{*}_{e3} U_{\mu 3} ] 
 \nonumber \\
 &=& \frac{1}{8} \sin 2 \theta_{12}  \sin 2 \theta_{23}  \sin 2 \theta_{13} 
 \cos \theta_{13}  \sin \delta,  
\ee 
where $\delta$ is the Dirac phase. 
The Jarlskog invariant as a function of the CP phase $\phi$ 
 is depicted in Fig.~7 
 for (a) the hierarchical and (b) the inverse-hierarchical cases, 
 respectively. 
We obtain a small but non-vanishing $J_{CP}$ 
 correlating with other outputs. 
In the hierarchical case, we find  
\be 
 J_{CP} = -4.8 \times 10^{-3}
\ee 
for $\phi=0.668$(rad).

\section{Conclusions} 

Neutrino oscillation experiments have explored neutrino masses 
 and mixing patterns. 
Tiny neutrino masses compared to the ordinary quark masses 
 are naturally explained by the see-saw mechanism 
 with heavy right handed neutrinos. 
Right handed neutrinos play the important role 
 to generate the baryon asymmetry in our universe 
 through leptogenesis. 
Leptogenesis requires CP violation in the lepton sector. 
For CP to violate we must have at least two right handed neutrinos. 
Keeping this minimal possibility in mind 
 we have introduced only two heavy right handed neutrinos 
 and studied a $3 \times 2$ Dirac type mass matrix $m_D$. 
Without loss of generality one can choose a reference basis 
 where both charged lepton mass matrix as well as the heavy right handed
 Majorana mass matrices are real and diagonal. 
In this basis, three physical CP phases appear in $m_D$. 

After the see-saw mechanism we obtain an effective $3 \times 3$ 
 Majorana mass matrix for light neutrinos. 
As a result from the $3 \times 2$ Dirac mass matrix $m_D$, 
 light neutrino mass spectrum should contain (at least) 
 one zero mass eigenvalue. 
This fact allows only two patterns for neutrino mass spectrum, 
 normal hierarchical or inverse-hierarchical.

We have chosen a simple texture for $m_D$ in our reference basis. 
To start with we have set all CP phases in $m_D$ to be zero. 
Although there is no CP violation in this case, 
 one can study this real texture in the context 
 of ongoing neutrino experiments. 
We have imposed an ansatz that the light neutrino mass matrix 
 is diagonalized by the tri-bimaximal mixing matrix. 
This ansatz strongly constrains model parameters, and 
 only four parameters 
 (two light neutrino  mass eigenvalues 
 and two heavy neutrino mass eigenvalues) 
 have been left free. 
Appropriate choice of two light neutrino mass eigenvalues 
 reproduces the current neutrino oscillation data.

Next, we have introduced a single CP phase in $m_D$. 
With four input parameters, we have examined 
 baryon asymmetry generated through leptogenesis 
 as well as neutrino oscillations, 
 as a function of only the CP phase. 
We can see interesting correlations between 
 resultant baryon asymmetry and neutrino oscillation parameters. 
For a special choice of the CP phase, 
 the hierarchical case can reproduce both the observed baryon 
 asymmetry and neutrino oscillation data. 
For a fixed CP phase reproducing the observed baryon asymmetry, 
 we have a prediction for a non-vanishing $|U_{e3}|$ 
 which is accessible in future baseline neutrino oscillation 
 experiments. 
In the inverse-hierarchical case, 
 we have not obtained sufficient baryon asymmetry 
 while resultant neutrino oscillation parameters 
 can be consistent with the current data.

Independently of the CP phase, 
 our texture leads to a unique relation for 
 the averaged neutrino mass relevant to 
 neutrino-less double beta decay experiments: 
 $m_{\nu_e \nu_e} = { \sqrt{\Delta m_{12}^2} \over 3}$ 
 and $ \sqrt{\Delta m_{23}^2} $ 
 in the hierarchical and the inverse-hierarchical cases,
 respectively. These results can be tested in next generation experiments 
 of neutrino-less double beta decay.   
We have worked in the context of a specific texture. However, as an
extension to our approach one can introduce small $c_1$ and check
whether the solutions reported in this article get drastically
modified. This is so because, often in real world models, one
may be able to restrict $c_1$ such that it is very small yet not
exactly zero. If a real $c_1$ is introduced in the complex case
($\phi \ne 0$), and its magnitude is of order $10 \%$ of the rest
of the $c_i$s, we see a $50\%$ variation in $\epsilon$ and $U_{e3}$.  
However, $\Delta m^2_{solar}, \Delta m^2_{atm}, \theta_{12}, \theta_{23}$
remain almost the same.

\section{Acknowledgments}

B.B is supported by UGC, New Delhi, India, under the grant number
F.PSU-075/05-06. He would also like to thank theory division KEK, Japan, for
hospitality and financial support where this work was performed.
The work of N.O is supported in part by Scientific Grants from
 the Ministry of Education and Science of Japan.


\newpage
\begin{figure}[ht, width=16cm, height=5cm]
\begin{center}
\subfigure[$\eta_B \times 10^{10}$ as a function of $\phi/\pi$ 
 in the hierarchical case]
{\includegraphics[height=4.8cm]{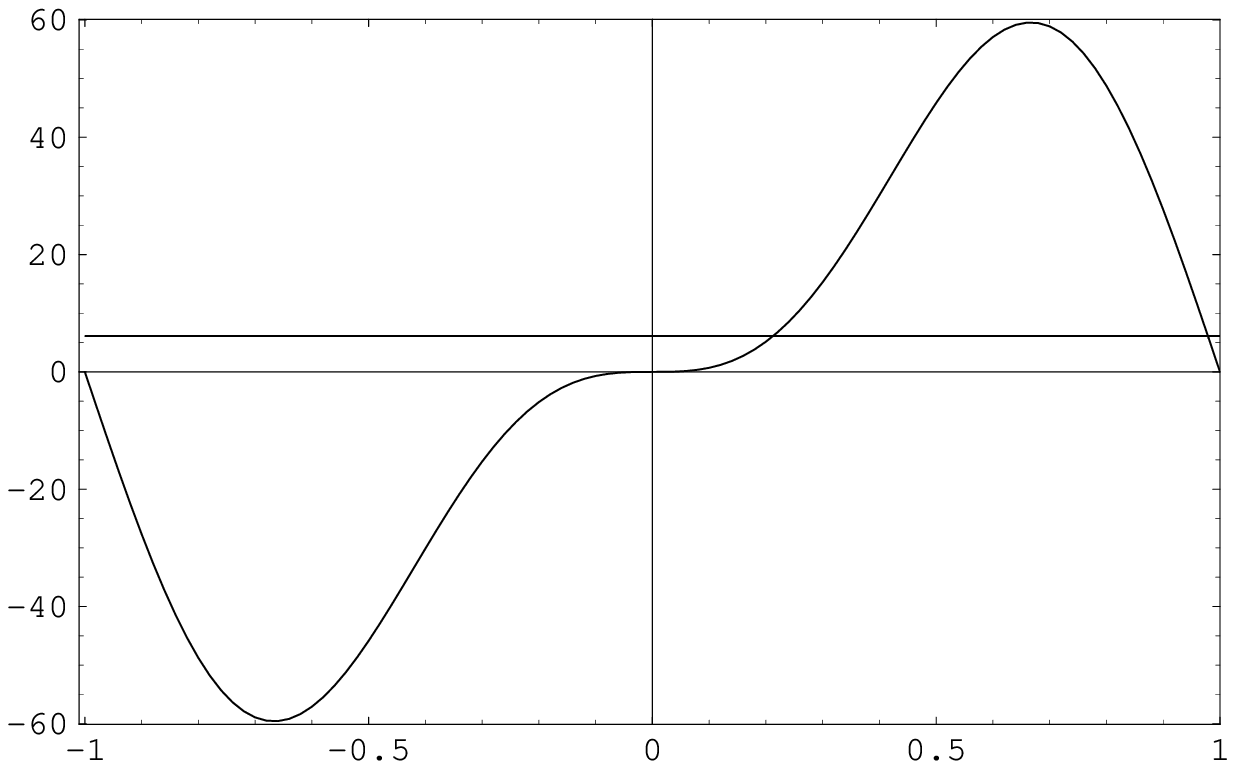}\label{Fig1a}}
\subfigure[$\eta_B \times 10^{10}$ as a function of $\phi/\pi$
 in the inverse-hierarchical case]
{\includegraphics[height=4.8cm]{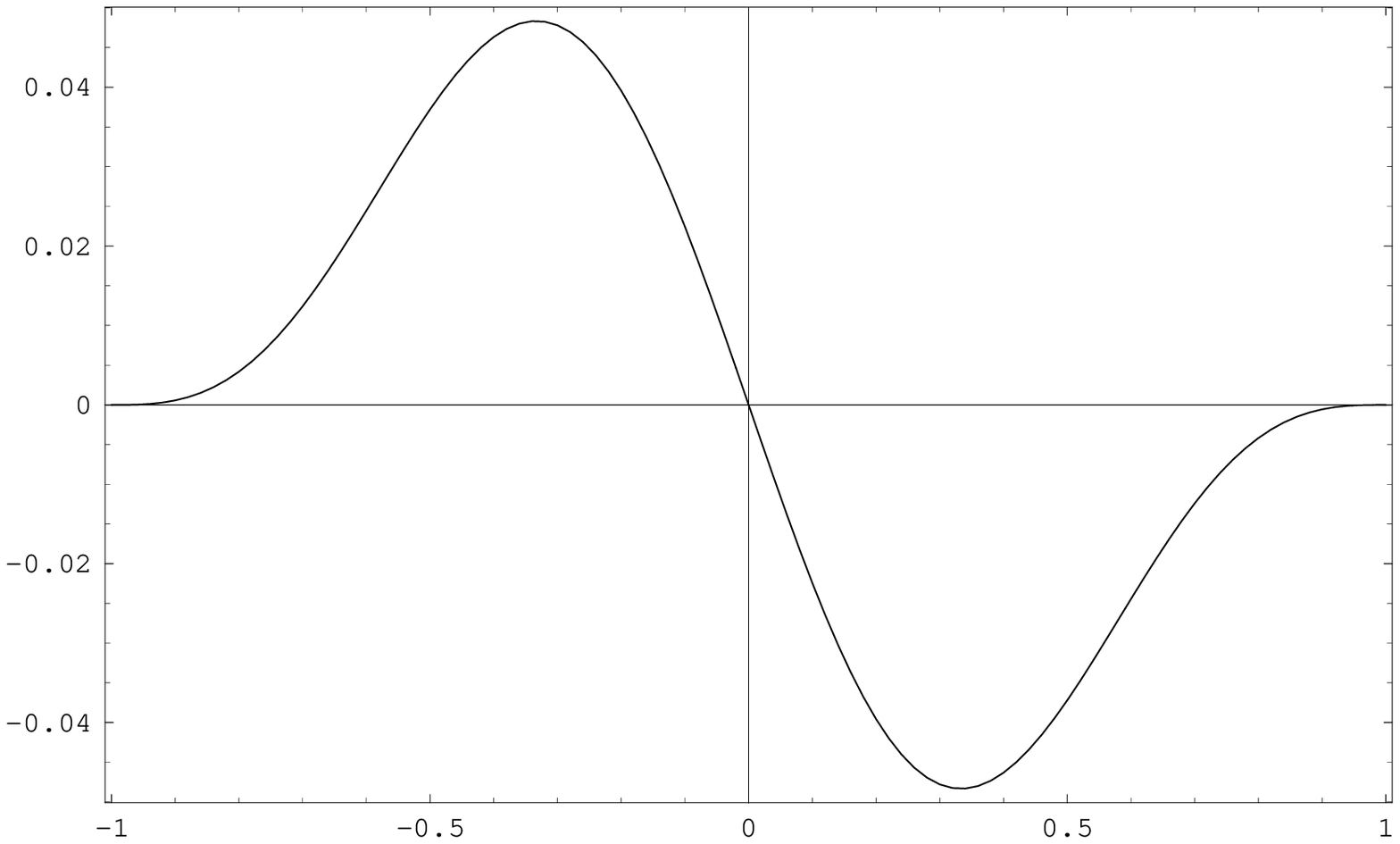}\label{Fig1b}}
\end{center}
\caption{
Baryon asymmetry as a function of the CP phase (in unites of $\pi$) 
 for (a) hierarchical case and (b) inverse-hierarchical case. 
We have used input values 
 $m_2= 9.59 \times 10^{-3}$ eV, 
 $m_3= 4.56 \times 10^{-2}$ eV, 
 $M_1=10^{13}$ GeV and $M_2=10^{14}$ GeV 
 for the hierarchical case, 
 while 
 $m_1= 4.46 \times 10^{-3}$ eV, 
 $m_2= 4.56 \times 10^{-2}$ eV, 
 $M_1=10^{13}$ GeV and $M_2=10^{14}$ GeV 
 for the inverse-hierarchical case. 
In Fig.~\ref{Fig1a}, the horizontal line corresponds to 
 the observed baryon asymmetry $\eta_B =6.1 \times 10^{-10}$. 
}
\end{figure}
%

\begin{figure}[hb, width=16cm, height=5cm]
\begin{center}
\subfigure[$ \Delta m_{12}^2 (\mbox{eV}^2)$ as a function of $\phi/\pi$]
{\includegraphics[height=4.8cm]{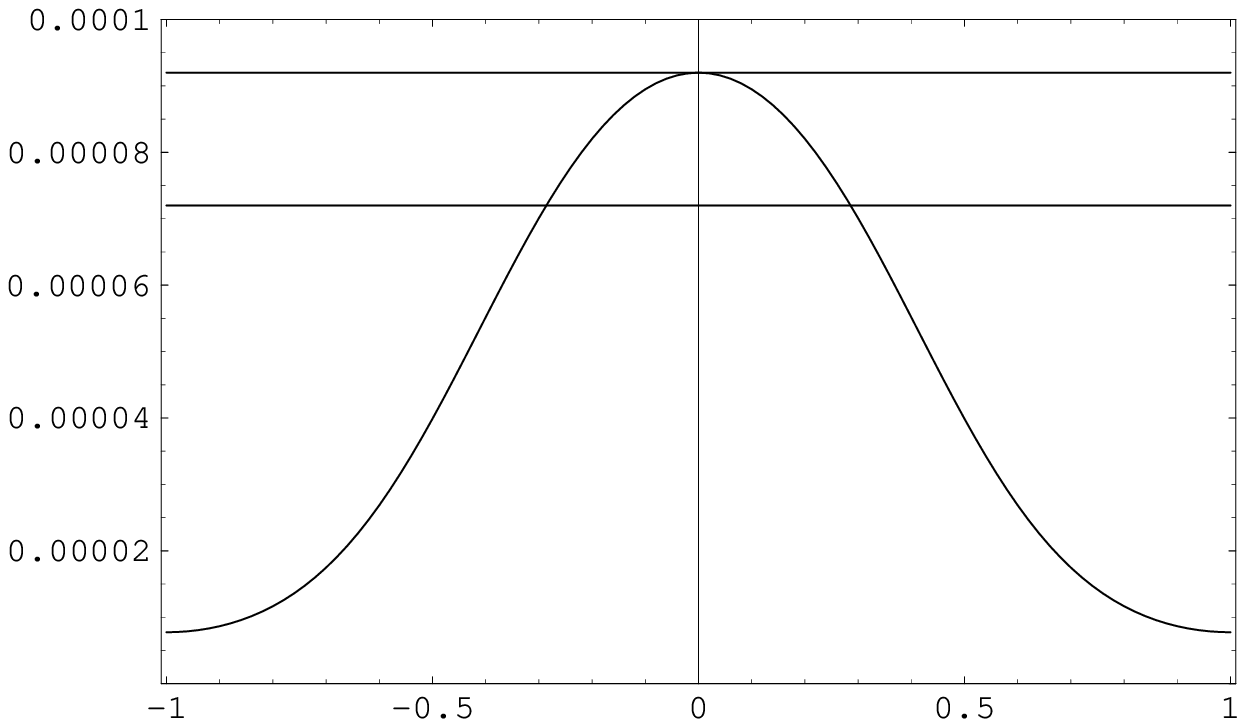}\label{Fig2a}}
\subfigure[$ \Delta m_{23}^2 (\mbox{eV}^2)$ as a function of $\phi/\pi$]
{\includegraphics[height=4.8cm]{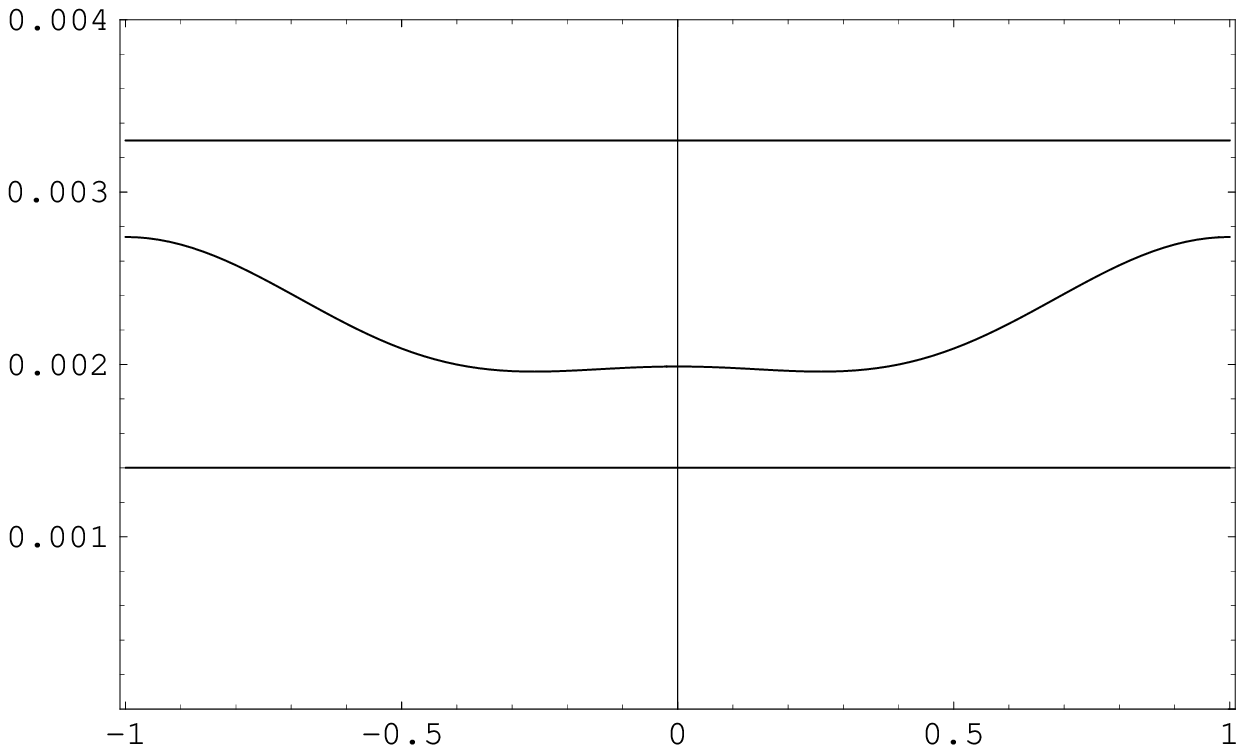}\label{Fig2b}}
\end{center}
\caption{
Mass squared differences as a function of the CP phase 
 (in unites of $\pi$). 
The region between two horizontal lines in each figure 
 are consistent with the current best fit values in 
 Eq.~(\ref{oscillation data}).
}
\end{figure}

\begin{figure}[H, width=16cm, height=6cm]
\begin{center}
\subfigure[$\sin^2 \theta_{12}$ as a function of $\phi/\pi$]
{\includegraphics[height=4.8cm]{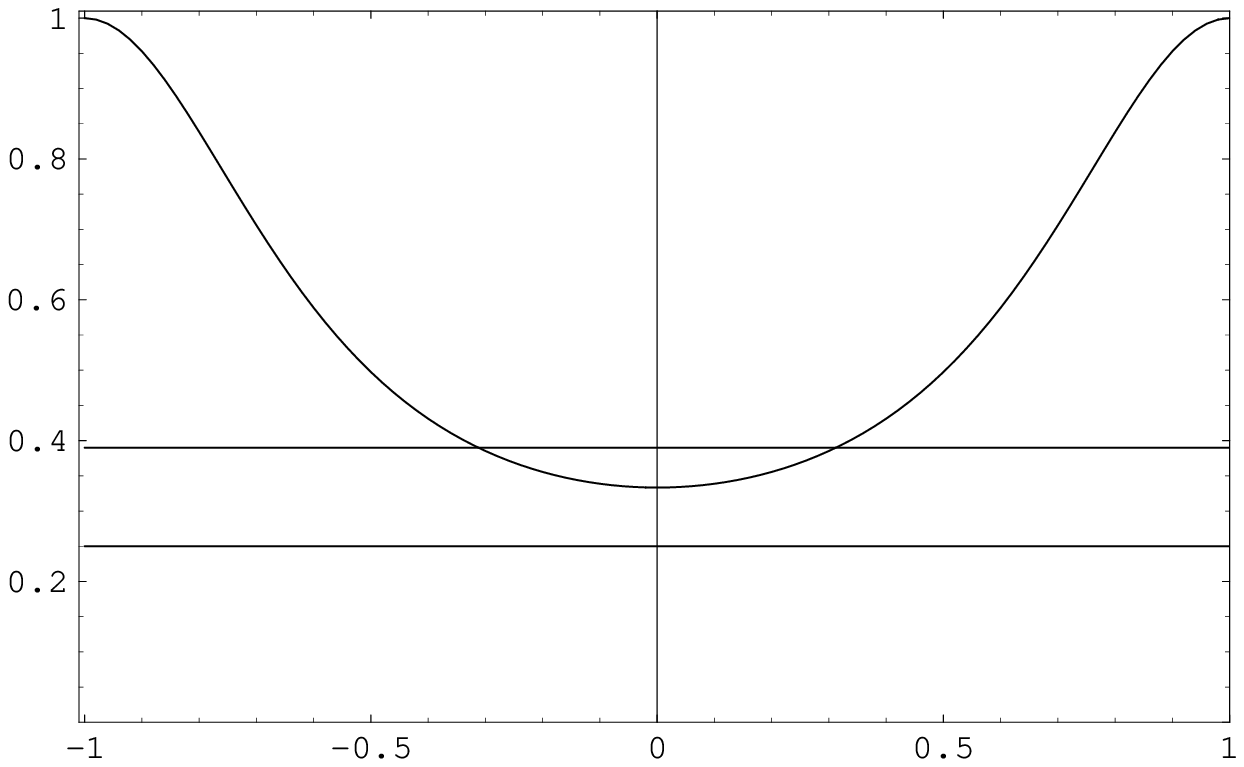}\label{Fig3a}}
\subfigure[$\sin^2 2 \theta_{23}$ as a function of $\phi/\pi$]
{\includegraphics[height=4.8cm]{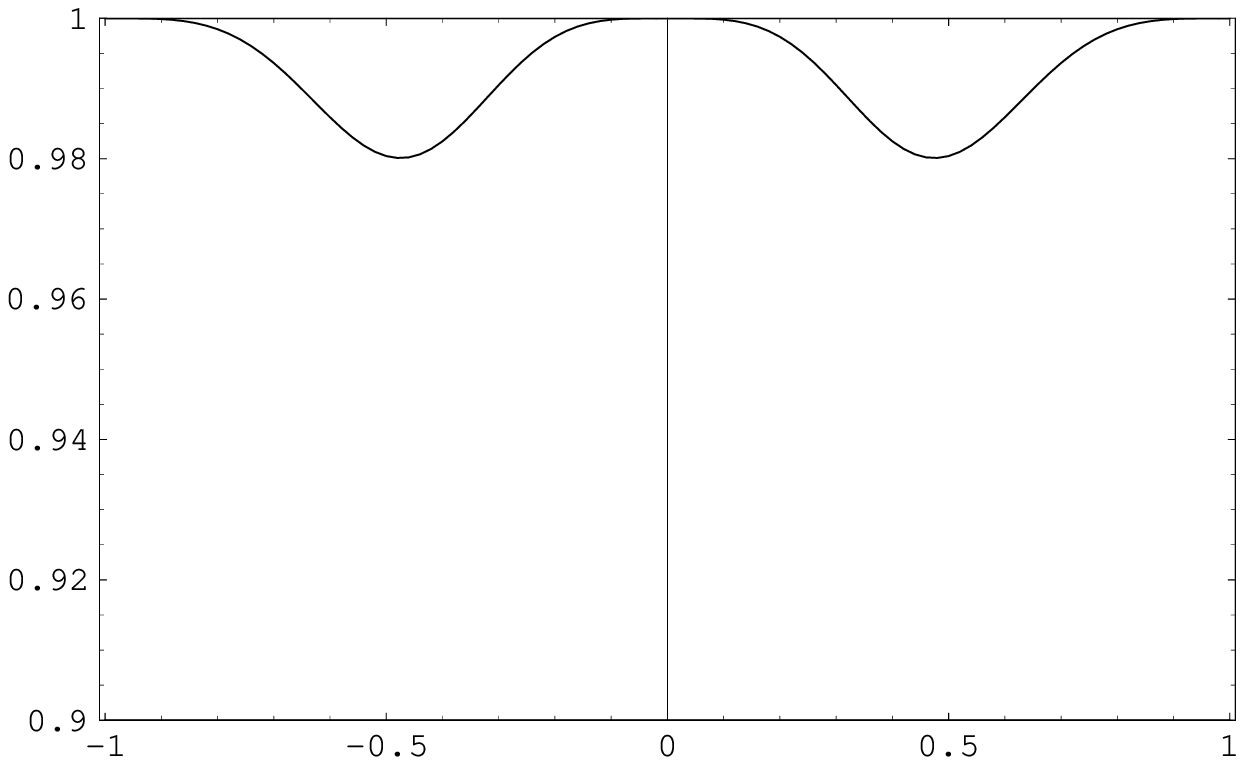}\label{Fig3b}}
\end{center}
\caption{
Neutrino mixing angles as a function of the CP phase 
 in unites of $\pi$. 
The region between two horizontal lines in Fig.~\ref{Fig3a} 
 is consistent with the current best fit values 
 in Eq.~(\ref{oscillation data}).
The entire region shown in Fig.~\ref{Fig3b} is allowed. 
}
\end{figure}

\begin{figure}[H, width=16cm, height=6cm]
\begin{center}
{\includegraphics[height=5cm]{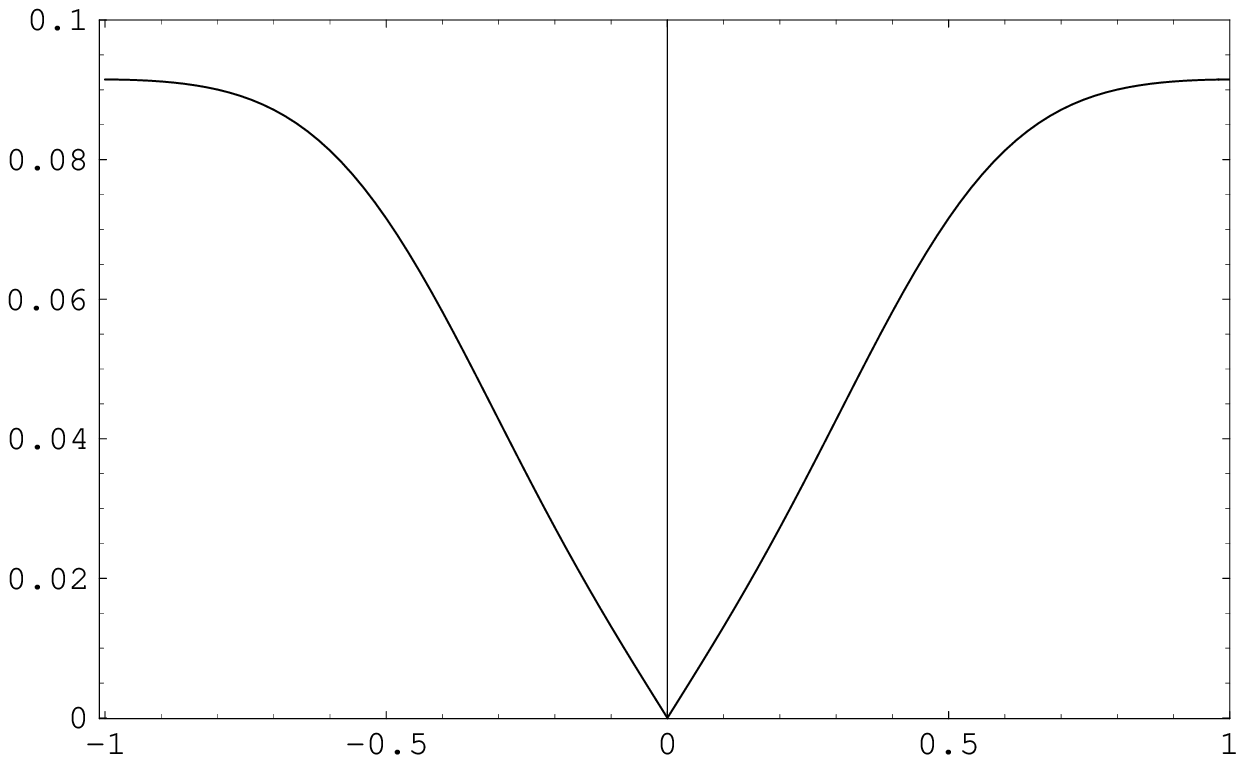}\label{Fig4}}
\caption{
$|U_{e3}|$ as a function of the CP phase in unites of $\pi$. 
}
\end{center}
\end{figure}

\begin{figure}[H, width=16cm, height=6cm]
\begin{center}
\subfigure[$ \Delta m_{12}^2 (\mbox{eV}^2)$ as a function of $\phi/\pi$]
{\includegraphics[height=4.8cm]{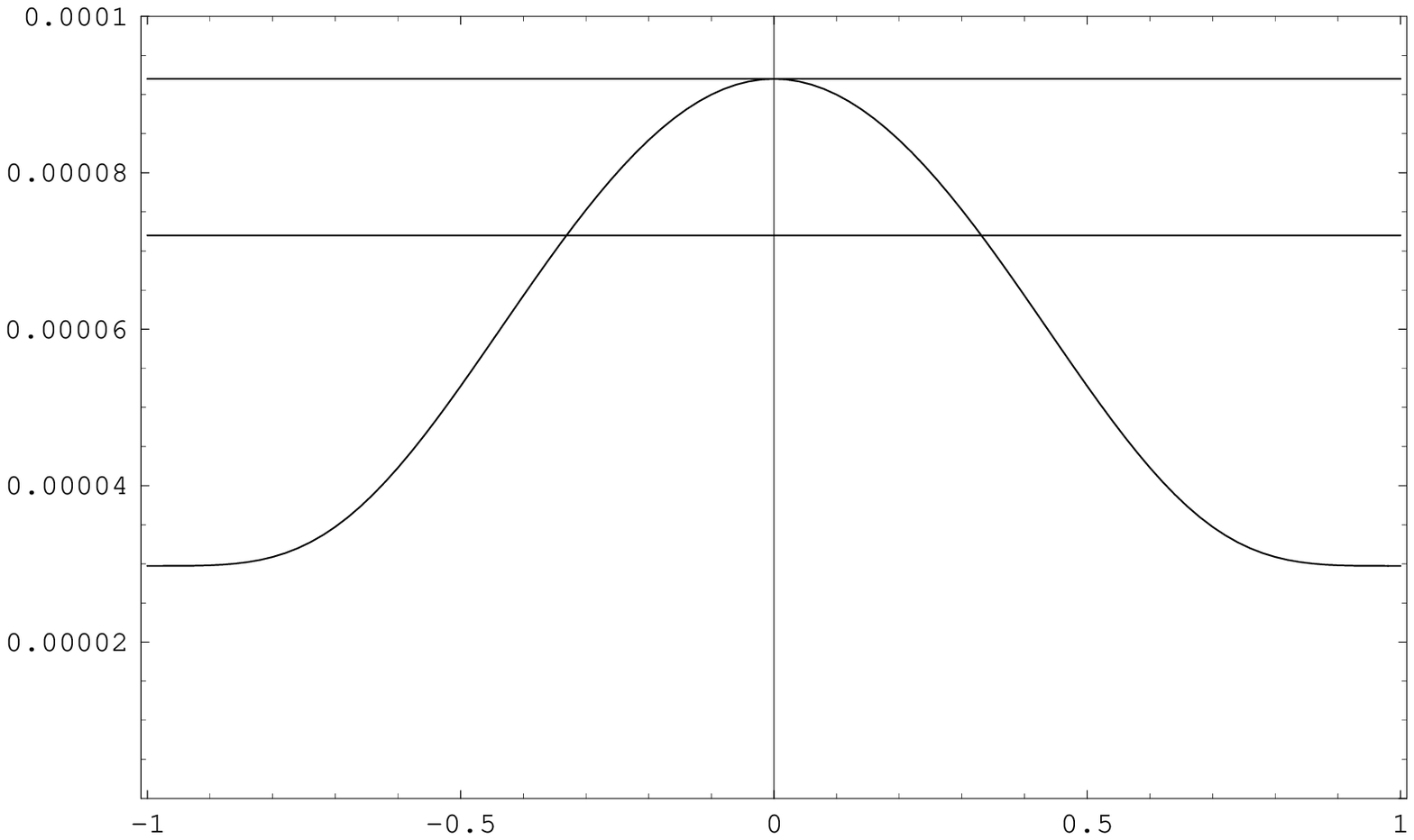}\label{Fig5a}}
\subfigure[$ \sin^2 \theta_{12}$ as a function of $\phi/\pi$]
{\includegraphics[height=4.8cm]{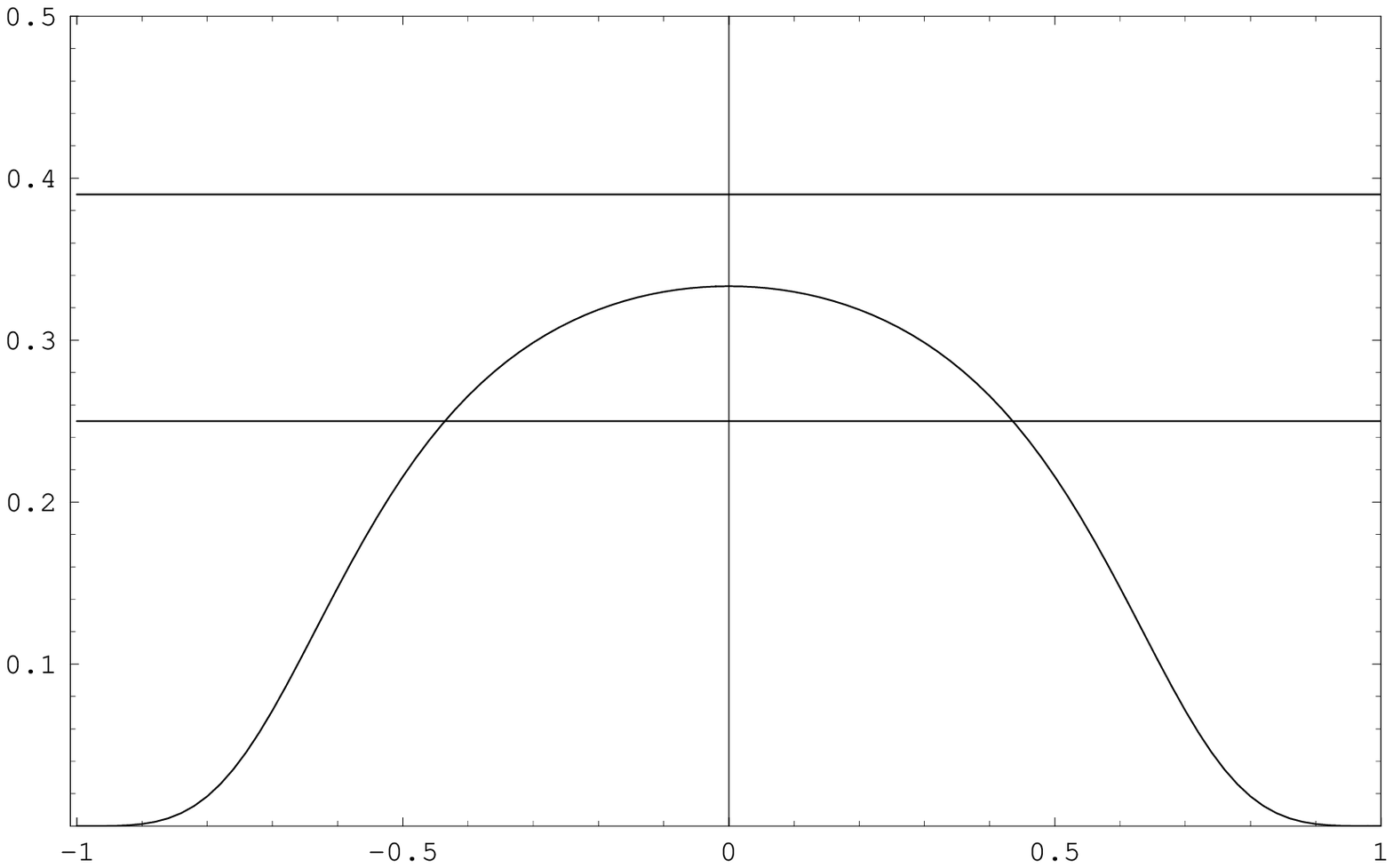}\label{Fig5b}}
\end{center}
\caption{
Mass squared difference and mixing angle 
 relevant for the solar neutrino oscillation 
 as a function of the CP phase in unites of $\pi$. 
The region between two horizontal lines in each figure 
 are consistent with the current best fit values in 
 Eq.~(\ref{oscillation data}).
}
\end{figure}

\begin{figure}[H, width=16cm, height=6cm]
\begin{center}
{\includegraphics[height=5cm]{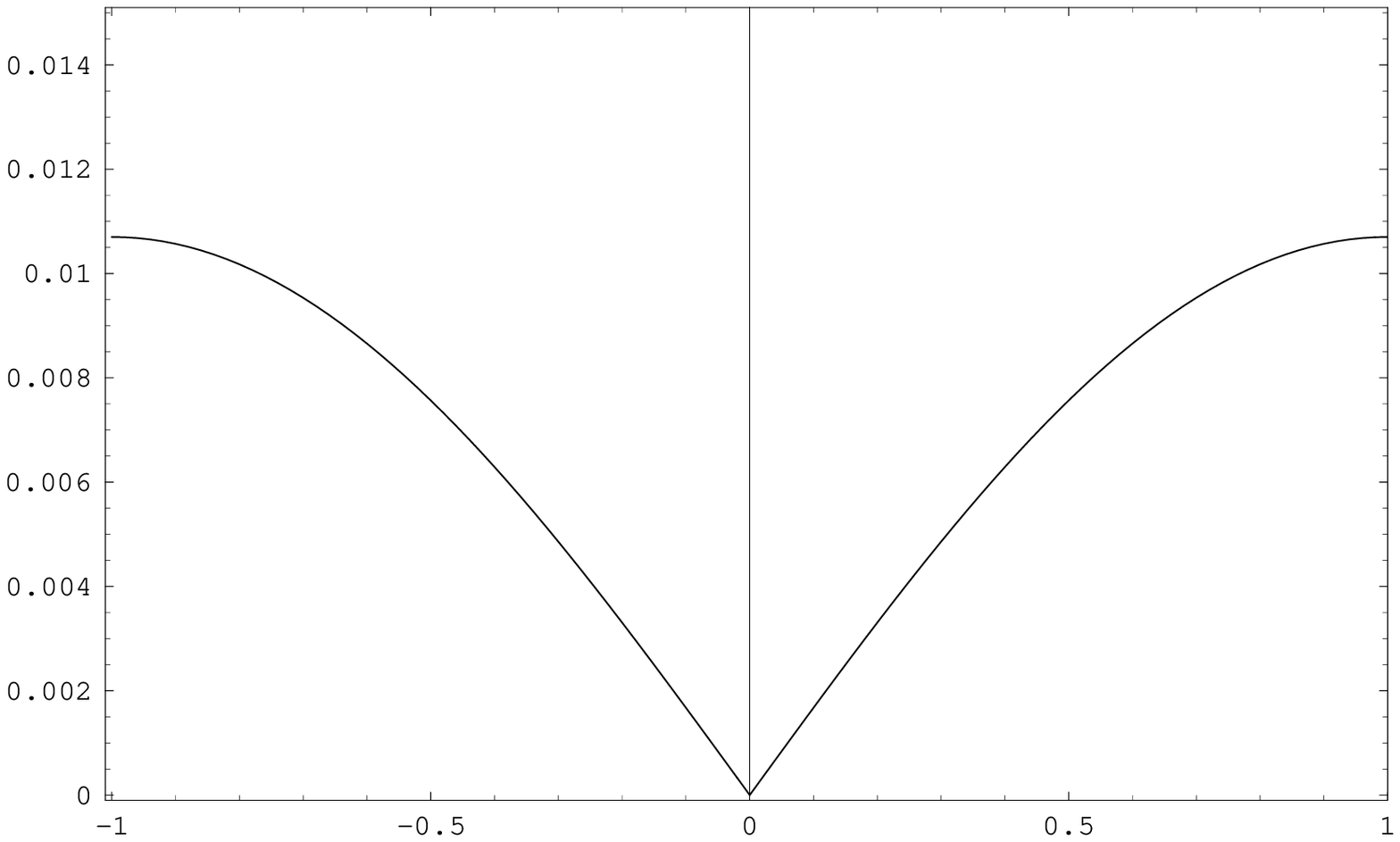}\label{Fig6}}
\caption{
$|U_{e3}|$ as a function of the CP phase in unites of $\pi$. 
}
\end{center}
\end{figure}

\begin{figure}[T, width=16cm,height=6cm]
\begin{center}
\subfigure[$J_{CP}$ as a function of $\phi/\pi$ 
in the hierarchical case]
{\includegraphics[height=4.8cm]{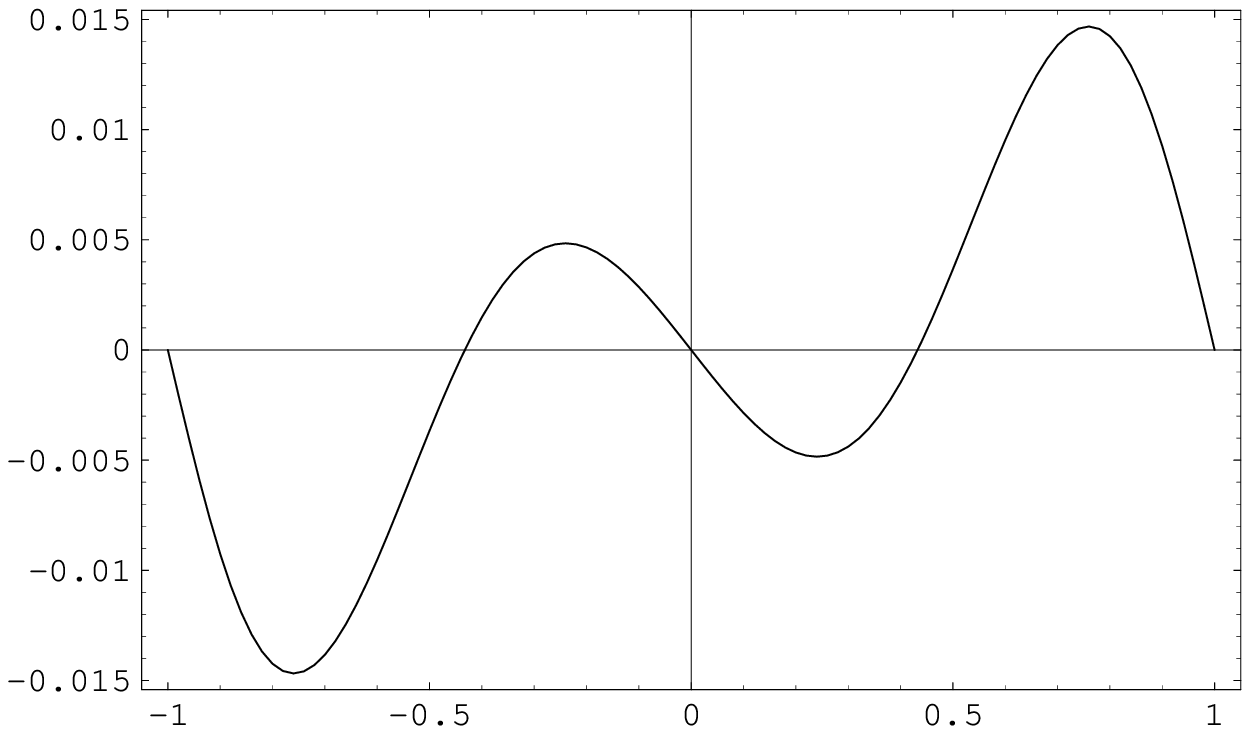}\label{Fig7a}}
\subfigure[$J_{CP}$ as a function of $\phi/\pi$
in the inverse-hierarchical case]
{\includegraphics[height=4.8cm]{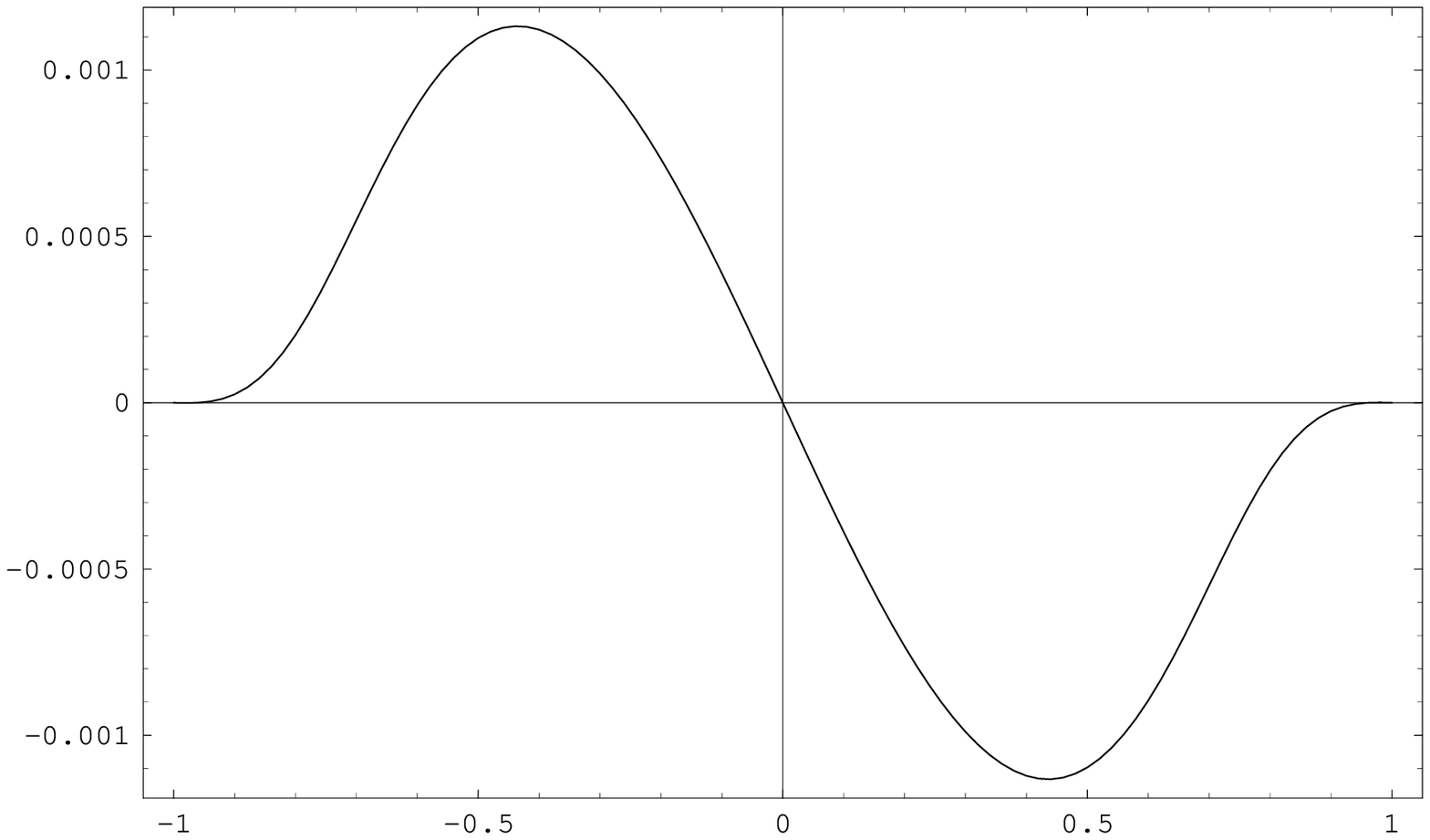}\label{Fig7b}}
\end{center}
\caption{
The Jarlskog parameter as a function of the CP phase (in unites of $\pi$) 
 for (a) hierarchical case and (b) inverse-hierarchical case. 
}
\end{figure}
%

\begin{thebibliography}{99}
 
\bibitem{WMAP-3Y}
D.N. Spergel {\it et al.}, arXiv:astro-ph/0603449.


\bibitem{bau}
S. Sarkar, Rept. Prog. Phys. {\bf 59} 1493 (1996);
R.H. Cyburt, Phys. Rev. {\bf D70} 023505 (2004). 

\bibitem{LG1}
M. Fukugita and T. Yanagida, Phys. Lett. {\bf B174} 45 (1986).  

%

\bibitem{LG2} 
For reviews see, 
W. Buchmuller and M. Plumacher, 
 Int. J. Mod. Phys. {\bf A15} 5047 (2000).

%
 

%


\bibitem{sph1}
V. A. Kuzmin, V. A. Rubakov, M. E. Shaposhnikov, Phys. Lett. 
{\bf B155} 36 (1985)


\bibitem{seesaw}
M. Gell-Mann, P. Ramond, and R. Slansky, in Supergravity, 
 edited by P. van Nieuwenhuizen and D. Z. Freedman 
 (North-Holland, Amsterdam, 1979), p. 315; 
T. Yanagida, in Proceedings of the Workshop on the Unified Theory 
 and the Baryon Number of the Universe, 
 edited by O. Sawoda and A. Sugamoto, KEK, 1979. 
R. N. Mohapatra and G. Senjanovic, 
 Phys. Rev. Lett. {\bf 44} 912 (1980);  
Phys. Rev. {\bf D23} 165 (1981);
J.W.F. Valle, in the proceedings of 8th Hellenic School on 
Elementary Particle Physics (CORFU2005), Corfu, Greece, 4-26 Sep 2005.
e-Print Archive: hep-ph/0608101. 


\bibitem{sakharov}
A. D. Sakharov, JETP Lett. {\bf 5} 24 (1967)


\bibitem{2rhn}
P.H. Frampton, S.L. Glashow and T. Yanagida, 
 Phys. Lett. {\bf B548} 119 (2002); 
A. Ibarra and G.G. Ross, 
 Phys. Lett. {\bf B591} 285 (2004); 
R. Gonzalez Felipe, F. R. Joaquim and B. M. Nobre, 
 Phys. Rev. {\bf D70} 085009 (2004);   
A. Ibarra, JHEP {\bf 0601} 064 (2006); 
K. Bhattacharya, N. Sahu, U. Sarkar and S.K. Singh, 
 e-Print Archive: hep-ph/0607272;   


\bibitem{nuosc1}
B. T. Cleveland {\it et.al}, Astrophys.J. {\bf 496} 505 (1998).


\bibitem{nuosc2}
Super-Kamiokande Collaboration, Phys. Lett. {\bf B539} 179 (2002); 
Super-Kamiokande Collaboration, Phys. Rev. {\bf D71} 112005 (2005).


\bibitem{nuosc3}
M. Maltoni, T. Schwetz, M.A. Tortola, J.W.F. Valle
 New J.Phys. {\bf 6} 122 (2004); 
A. Bandyopadhyay {\it et al}, 
 Phys. Lett. {\bf B608} 115 (2005); 
G. L. Fogli {\it et al}, 
 Prog. Part. Nucl. Phys. {\bf 57} 742 (2006).


\bibitem{hps}
P. F. Harrison, D. H. Perkins, W. G. Scott, Phys. Lett. {\bf B530} 167 (2002) 


%
%

\bibitem{vissani}
M. Flanz, E. A. Paschos and U. Sarkar, 
 Phys. Lett. {\bf B345} 248 (1995); Erratum-ibid. {\bf B382} 447 (1996); 
L. Covi, E. Roulet and F. Vissani, 
 Phys. Lett. {\bf B384} 169 (1996);  
W. Buchmuller and M. Plumacher, 
 Phys. Lett. {\bf B389} 73 (1996); 
 ibid. {\bf B431} 354 (1998). 


\bibitem{lgn1}
M.A. Luty, 
 Phys. Rev. {\bf D45} 455 (1992); 
M. Plumacher, 
 Z. Phys. {\bf C74} 549 (1997); 
 Nucl. Phys. {\bf B530} 207 (2002); 
R. Barbieri, P. Creminelli, A. Strumia and N. Tetradis, 
 Nucl. Phys. {\bf B575} 61 (2000); 
W. Buchmuller, P. Di Bari, M. Plumacher, 
 Nucl. Phys. {\bf B643} 367 (2002).  


\bibitem{lgn2}
W. Buchmuller, P. Di Bari and M. Plumacher,
 Annals Phys.  {\bf 315}, 305 (2005); In the
context of tri-bimaximal mixing see,  Zhi-zhong Xing and Shun Zhou,
hep-ph/0607302.




\bibitem{mike}
W. Buchmuller, M. Plumacher, Phys. Rept. {\bf 320} 329 (1999); 
W. Buchmuller, T. Yanagida, Phys. Lett. {\bf B445} 399 (1999); 
M. S. Berger, B. Brahmachari, Phys. Rev. {\bf D60} 
073009 (1999);  
D. Falcone, F. Tramontano, Phys. Rev. {\bf D63} 073007
(2001);  
D. Falcone, F. Tramontano, Phys. Lett. {\bf B506} 1 (2001)



\bibitem{rlg}
M. Flanz, E.A. Paschos, U. Sarkar and J. Weiss,
 Phys. Lett. {\bf B389}, 693 (1996); 
L. Covi and E. Roulet,
 Phys. Lett. {\bf B399}, 113 (1997); 
A. Pilaftsis,
 Nucl. Phys. {\bf B504}, 61 (1997); 
 Phys. Rev. {\bf D 56}, 5431 (1997). 


\bibitem{PPR}
S. Pascoli, S.T. Petcov and A. Riotto,
arXiv:hep-ph/0611338; G.C. Branco, R. Gonzalez Felipe, F.R. Joaquim
e-Print Archive: hep-ph/0609297; S. Antusch, S.F. King, A. Riotto, 
JCAP {\bf 0611}, 011 (2006)


\bibitem{Jarlskog}
C. Jarlskog,
 Phys. Rev. Lett. {\bf 55}, 1039 (1985).


\end{thebibliography}
\end{document}